\documentclass[12pt,a4paper]{article}
\usepackage{amsmath, amsfonts, amssymb, epsfig, palatino, fancyhdr}
\begin{document}

\begin{titlepage}

\begin{center}

{\Large \bf Loschmidt's paradox, entropy and the topology of spacetime }

\vspace{15mm}

{\bf Moninder Singh Modgil \footnote[1]{PhD, Physics from Indian Institute of Technology, Kanpur, India; B.Tech. (Hons.) from Indian Institute of Technology, Kharagpur, India \\ email: moni\_g4@yahoo.com}  }

\vspace{1cm}

{\bf Abstract}

\end{center}

The issue of the "eternal return" is examined from  the perspective of the topology of spacetime. Constraints on dynamical laws for the periodic evolution of a system or universe are highlighted. In a periodic universe with period $T$, using a Fourier series expansion for any physical variable $A(t)$ and its time derivatives $A^n(t)$, an infinite set of simultaneous linear equations is formulated. 
\vspace{5mm}

\noindent {\bf PACS:} 05.70.-a, 05.90.+m, 04.20.Jb, 04.20.Gz 

\vspace{5mm}
\noindent {\bf KEY  WORDS:}  Entropy, eternal return, Loschmidt's paradox, time periodic universe

\end{titlepage} 

Frank Tipler \cite{Tipler 1979, Tipler 1980}  proved a no-return theorem for general relativity. Loschmidt's paradox \cite{Loschmidt 1876} states that an irrversible process, such as the monotonically increasing entropy function, cannot arise from  time symmetric dynamics. It was formulated by Loschmidt to counter the derivation of second law of thermodynamics by Boltzmann using the Boltzmann equation. The essential argument by Loschmidt is that a \textit{velocity reversion},

\begin{eqnarray}
	v \rightarrow -v
\end{eqnarray}
\noindent for all the particles constituting the system, leads to recurrence in the time evolution of the system. 
 
To see that spacetime topology is a player in the eternal return problem, consider the following example of a Recurrence metric \cite{Modgil 2001}, periodic over $t \in [-T,T]$. This universe is a vacuum solution of the Einstein's field equation. In this universe, the  velocity reversion for all particle of the universe occurs due to the compact topology of time, at the time instants $t=-T$, and $t=T$; which respectively are the past and future boundary, of this spacetime. The topology of spacetime  is $R^3 \times S^1/Z_2$, ($R^3$ - space, and $S^1/Z_2$ - time), and the metric  is given by the line element-

\begin{eqnarray}\label{Recurrence Metric}
	ds^2 = \left(\frac{\pi}{T} \right)^2 \cos^2 \frac{ \pi t}{T} dt^2 -dx_1^2- dx_2^2-dx_3^2 
\end{eqnarray}

\noindent 

Integration of the geodesic equations of this Recurrence metric, yields sinusoidal particle paths,with period $T=2 \pi$ -

\begin{eqnarray}
\begin{array}{cc}
	x_i=v_i(0) \sin x_0+x_i(0), & i=1,2,3
\end{array}
\end{eqnarray}

\noindent where $v_i(0)$ and $x_i(0)$ are the initial velocity and position of the particles, at $x_0 = 0$. The universe described is recurring for non-interacting particles. For point inter-particle interaction, the particle paths are piece wise geodesic - as before and after the collisions, the particles are moving on geodesics. At the instants of velocity reversion the particles simply reverse their direction of motion - experience the same chain of collisions in the reverse direction - and hence ensure recurrence.

By defining the entropy function $S(t)$, 
\begin{eqnarray}
	S(t) \sim \sin \frac{\pi t}{T}
\end{eqnarray}

\noindent the reversal of arrow of time, i.e., 

\begin{eqnarray}
	\frac{dS}{dt}=0
\end{eqnarray}
\noindent at the past and future temporal boundaries ($t=-T/2$, and $t=T/2$ respectively) becomes clear. Thus there exist two phases of the universe's evolution - 

\begin{enumerate}
\item
(1) During which the entropy increases, 

\begin{eqnarray}
\begin{array}{cc}
	dS(t)/dt \geq 0 & t \in [-T/2, T/2]
\end{array}
\end{eqnarray}

and,

\item During which the entropy decreses,

\begin{eqnarray}
\begin{array}{cccc}
	dS(t)/dt \leq 0, & t \in (T,-T/2), & and & t \in [T/2,T)
\end{array}
\end{eqnarray}
 
\end{enumerate}
 
 The sigularities at joining points of these two phases, follow trivially from the Morse theory \cite{Milnor 1963}. Morse theory  requires that for any function $f$ defined on a closed curve $\gamma$, parameterized by $s$, there exists an even number of  critical points, on which the derivative of the function vanishes, i.e., $df/ds = 0$. 

Note that compact topology of the time coordinate is also encountered in - (1) Conformal compactification of Minkowski spacetime -

\begin{eqnarray}
	R^3 \times R^1 \rightarrow S^3 \times S^1
\end{eqnarray}

\noindent and (2) in the Anti de Sitter spacetime. However, these metrics, lack an explicit mechanism to ensure recurrence of particle positions and velocities.

One may note the following subtle distinctions, for recurrence in a  cosmological model with $S^1$ topology of time -

\begin{enumerate}

\item $S^1$ topology of time with a finite period $T$, is remniscient of attempts by Poinc$\acute{a}$re \cite{Brush 1966},  to prove eternal recurrences in an eternal universe. However, a distiction has to be made between the "exact recurrence", in a finite time implied by the $S^1$ topology of time, and the concept of "asymptotic Poinc$\acute{a}$re recurrence", as used in ergodic theory \cite{Barreira 2006}. The Poinc$\acute{a}$re recurrence theorem states that certain systems will, after a sufficiently long time, return to a state very close to the initial state. Naively, the asymptotic Poinc$\acute{a}$re recurrence can be considered equal to the case of exact recurrence in $S^1$ topology of time, provided the time periodicity $T$ of recurrence in $S^1$ time, is allowed to approach infinity - as now the system or the universe has sufficient time to return to the initial state. Recurrence metric eq.(\ref{Recurrence Metric}) in the limit $T \rightarrow \infty$  mirrors asymptotic periodic evolution of the universe in the interval $[-\infty, \infty]$ with identical initial and final condition.

\item Periodic cosmology of $S^1$ time is in general distinct from the Friedmann-Robertson-Walker (FRW) oscillating cosmological model with positive curvature $(k=1)$ \cite{Jaki 1974}. Using definition of wave function of universe, Hawking \cite{Hawking 1985} was able to connect thermodynamic arrow of time with the cosmological arrow, i.e., entropy decreases during the collapsing phase. In an accompanying paper, Page \cite{Page 1985}, argued that classically observable individual WKB components will not have their thermodynamic arrow reversed in the collapsing phase.  Remarkably, Cocke \cite{Cocke 1967} argues that entropy can decrease in the collapsing phase of this universe without microscopic motion reversal!

\item Consider the winding of the $R^1$ universal cover over the  $S^1$ time. In this universal cover, each event has an infinite numbers of copies in past and future. Where as in $S^1$ time, each event is unique, i.e., there are no replications. The initial concept of the {\it eternal return} of Poinc$\acute{a}$re therefore, more accurately corresponds to this universal cover.

\end{enumerate}

$S^1$ topology of time requires periodic boundary conditions for all physical variables, which in turn evolve according to the dynamical laws. This in turn imposes constraints on the dynamical laws themselves, operating in such a universe! Indeed, formulation of such dynamical laws is in author's view, the preliminary research problem which $S^1$ topology of time generates. Following Popper's \cite{Popper 1957} criteria of {\it falsifiability of scientific theories}, whether such dynamical laws are consistent with what are observed in nature, is the touchstone for this topology of time.

 If $A$ is any variable, and $A^n$ its $n$-th derivative with respect to the time $t \in [0,T]$ then, the dynamical laws operative in the time periodic universe with period $T$, should be such that -  

\begin{eqnarray}
\begin{array}{ccc}
	\int_{0}^{T} A^n(t) dt = 0, & for & n=1,2,3,...\infty
	\end{array}
\end{eqnarray}

\noindent The change in value of $A$, over any time interval $(t_1,t_2)$ is required to be negative of the total change during the rest of the cycle, i.e., as -

\begin{eqnarray}
	\int_{0}^{t_1} A^1(t) dt +  \int_{t_1}^{t_2} A^1(t) dt +\int_{t_2}^{T} A^1(t) dt = 0
\end{eqnarray}

\noindent hence,

\begin{eqnarray}
	\int_{t_1}^{t_2} A^1(t) dt = - \left(  \int_{0}^{t_1} A^1(t) dt + \int_{t_2}^{T} A^1(t) dt \right)
	\end{eqnarray}

\noindent It follows that $\delta A$, the instantaneous change in $A(t)$, at the time instant $t=\tau$, is the opposite of the change during the rest of the time cycle -

\begin{eqnarray}
	\delta A |_{t=\tau} =  \lim_{\epsilon \rightarrow 0} \left[ A(\tau+\epsilon)-A(\tau-\epsilon) \right]= - \left(  \int_{0}^{\tau-\epsilon} A^1(t) dt + \int_{\tau + \epsilon}^{T} A^1(t) dt \right)
\end{eqnarray}
\noindent where, $\epsilon$ is an infinitesimal time interval.

Now, lets apply Morse theory to $A(t)$ and its derivatives $A^n(t)$.  Let $A^n$ vanish at the ordered set of points $X^n=\{t_1^n, t_2^n, t_3^n,\ldots\}$ consisting of an even number of elements, with $t_1^n < t_2^n, < t_3^n,\ldots$. On the interval  $[t_i^n, t_{i+1}^n]$, one can define a periodic function $B^n = A^n(t)$, with the periodicity $T_{B^n} = (t_{i+1}^n - t_i^n)$. Applying Morse theory again for $B^n$ (periodic over the interval $T_{B^n}$) we have the straight forward result that $B^{n+1} = A^{n+1}$ vanishes over the ordered set $X^{n+1} = \{t_1^{n+1}, t_2^{n+1}, t_3^{n+1},\ldots \}$ which consists of an even number of elements, with $t_1^n < t_1^{n+1} < t_2^{n+1} < t_3^{n+1},...$. This result holds for all $n$. Apparently, we have a fractal structure for distribution of vanishing points of derivatives of $A(t)$. 

Now, consider the simplified picture of a time periodic universe consisting of $N$ particles. Let $p$ and $q$ be the generalized momenta and coordinates respectively, with $p=\{p_1, p_2, p_3, \ldots, p_N\}$ and $q=\{q_1, q_2, q_3, \ldots, q_N\}$, where $p_i$ and $q_i$ are momentum and coordinates of the $i$-th particle. For reccurrence of the initial conditions, it is required that trajectory $\gamma$ of the universe in the phase space be closed. We have,
\begin{eqnarray}
	\oint_{\gamma} dp = 0
	\oint_\gamma dq = 0
\end{eqnarray}

A distinction between the two recurrence conditions has to be made - 

\begin{enumerate}

\item Identical initial and final condition, 

\begin{eqnarray}
	q(0)=q(T),
\end{eqnarray}
 
 \item Identical evolution across all time cycles, 
  
\begin{eqnarray}
	q(t) = q(t+T)
\end{eqnarray}
  
 \end{enumerate}
 
 \noindent  This distinction is readily seen as follows. Note that mere repetition of position $q(0)=q(T)$ and momenta $p(0)=p(T)$ over the time interval $t \in [0,T]$, by itself does not ensure that the evolution in each cycle will be identical, i.e.,  if the accelerations $a(t) = dp(t)/dt$ are different,  ($a(0) \neq a(T)$),   the evolution in subsequent cycle $t \in [T, 2T]$ will be different. In other words, trajectory in the phase space of the system in the two cycles will be different. Similarly, merely insuring repetition of acceleration, along with position and velocity, again is not an insurance for identical evolution, because the jerk $j=da(t)/dt$ can be different ($j(0) \neq j(T)$. Hence, while the condition $q(0) = q(T)$, ensures recurrence of the initial conditions; the condition $q(t) = q(t+T)$, assures identical evolution across the time cycles. This identical evolution across time cycles may be achieved by - (1) the existence of all derivaties $q^n(t)=d^n q(t)/dt^n$, and (2) the recurrence in all derivatives, namely $q^n(t)=q^n(t+T)$.

Now, consider the expansion of any physical variable $A$, (e.g., position, energy, etc., of a particle or even a set of N particles) in a Fourier series -

\begin{eqnarray}\label{Fourier Expansion}
	A(t) = \sum_{m=0}^{m=\infty} \left( C_m^1 \cos \frac{2 \pi m t}{T}+ C_m^2 \sin \frac{2 \pi m t}{T} \right)
\end{eqnarray}

\noindent Expression for the derivatives $A^n(t$), can be obtained by differentiating equation (\ref{Fourier Expansion}), $n$ times, with respect to $t$. Thus, given an infinite set of the initial conditions, $A(0)$ and $A^n(0)$, one obtains an infinite set of simultaneous linear equations -
 
 \begin{eqnarray}
 \begin{array}{cc}
\nonumber	 A(0) = \sum_{m=0}^{m=\infty} C_m^1, &
\\ \nonumber  A^n(0) = \sum_{m=0}^{m=\infty} \left( \frac{2 \pi m}{T} \right)^m C_m^2, &  n \in Z^{Odd}-3 Z^+, 
\\  \nonumber  A^n(0) = \sum_{m=0}^{m=\infty} -\left( \frac{2 \pi m}{T} \right)^m C_m^1, & n \in Z^{Even} -4 Z^+ , 
\\ 	\nonumber  A^n(0) = \sum_{m=0}^{m=\infty} - \left( \frac{2 \pi m}{T} \right)^m C_m^2, & n \in 3 Z^+, 
\\ 	 A^n(0) = \sum_{m=0}^{m=\infty} \left( \frac{2 \pi m}{T} \right)^m C_m^2, & n \in 4 Z^+  
	\end{array}
\end{eqnarray}

\noindent Here $Z^+,Z^{+Even}, Z^{+Odd}$ are set of positive integers, positive even integers and positive odd integers respectively; $n Z^+$ represents the set of those positive integers, which are multiples of the integer $n$; and $(-)$ represents the operation of set complement, i.e., $X-Y$ means, the set of those elements of $X$, which are not there in $Y$. 

Dienes \cite{Dienes 1932}, discusses such systems of infinite linear equations, 

\begin{eqnarray}\label{Infinite Matrix}
\begin{array}{cc}
	\sum_{m=1}^{\infty} a_{nm} x_m = y_n, & n=1,2,\ldots, \infty
	\end{array}
\end{eqnarray}

\noindent for which, for a given $y_n$, the solution for $x_n$ depends on matrix characteristics of the infinite matrix $a_{nm}$. Cooke \cite{Cooke 1955}, Shivakumara and Sivakumar \cite{Sivakumar 2009} describe properties of such infinite matrices and sequence spaces. One of the initial applications of inifnite matrices was in the Heisenberg-Dirac theory of quantum mechanics.  Infinite matrices, are the forerunner and a main constituent of many branches of classical mathematics (infinite quadratic forms, integral equations, differential equations, etc.) and of the modern operator theory.

Since,

\begin{eqnarray}
	A^n(t) \propto \frac{1}{T^n},
\end{eqnarray}

\noindent therefore in the limit $T \rightarrow \infty$

\begin{eqnarray}
\begin{array}{cc}
	\lim_{t \rightarrow \infty} A^n(t) = 0, & n=1,2,... \infty
\end{array}
\end{eqnarray}

\noindent which implies a universe with no dynamics, or a {\it frozen universe}. 

 From practical considerations, an experimental measurement of a physical observable $A(t)$, lasting over a finite time interval $\Delta t$, will not provide sufficient information to determine all its Fourier modes. In general, one can impose an Infra-red (IR), cut off $m_{IR}$ and an Ultra-Violet (UV) cut off  $m_{UV}$ for the {\it measurable} Fourier modes, due to various reasons - e.g. (1) finite resolution of the of experimental apparatus, especially the clocks, (2) the quantum mechanical nature of the system/universe, (3) In the particle position, UV cutoff can arise due to a discrete spacetime \cite{Snyder 1947}, while the IR cutoff can arise due to $S^3$ topology of the space, i.e., a finite size of the universe. For such a measurement the matrix $a_{nk}$, of equation (\ref{Infinite Matrix}), is replaced by a finite one.  
 
 \noindent {\bf Acknowledgement}
 \\
 I would like to thank John Barrow for drawing my attention to Tipler's \cite{Tipler 1980} work.


\begin{thebibliography}{unsrt}

\bibitem{Tipler 1979} Tipler, F.:  \textit{Nature} \textbf{280}, 203-5 (1979).

\bibitem{Tipler 1980} Tipler, F.: \textit{Essays in general relativity: festschrift for A Taub}, ed F. J. Tipler, Academic press pp21-37 (1980).

\bibitem{Loschmidt 1876}  Loschmidt, Sitzungsber. Kais. Akad. Wiss. Wien, Math. Naturwiss. Classe \textbf{73}, 128–142 (1876).

\bibitem{Modgil 2001} Modgil, M.S., and Sahdev, D.: \textit{Recurrence metrics and the physics of closed time-like curves}, (2001), arXiv:gr-qc/0107055

\bibitem{Milnor 1963} Milnor, J.:{\it  Morse Theory}, (1963), Princeton University Press.

\bibitem{Brush 1966} Brush, Stephen G.: \textit{Kinetic Theory, Vol.2, Irreversible processes}, Pergamon Press,
(1966).

\bibitem{Barreira 2006} Barreira, L.: {\it Poinc$\acute{a}$re recurrence: old and new}, in Zambrini, Jean-Claude, XIVth International Congress on Mathematical Physics, , pp. 415–422, (2006), World Scientific, pp. 415–422,

\bibitem{Jaki 1974} Jaki,S.L.: \textit{ and Creation- from eternal cycles to
oscillating universe}, Scottish Academy Press, (1974).

\bibitem{Hawking 1985} Hawking, S.W.: {\it Arrow of time in cosmology}, Phys. Rev. D, {bf 32}, 2489, (1985).

\bibitem{Page 1985} Page, D.: {\it Will entropy decrease if the Universe recollapses?}, Phys. Rev. D, {bf 32}, 2496, (1985).

\bibitem{Cocke 1967} Cocke, W.J.: {\it Statistical Time Symmetry and Two-Time Boundary Conditions in Physics and Cosmology}, Phys. Rev. D, {\bf 160}, 1165, (1967).

\bibitem{Popper 1957} Popper, K.R.: \textit{Quantum theory and the schism in physics}, (1957), Hutchinson.

\bibitem{Dienes 1932} Dienes, P.: {\it Notes on Linear Equations in Infinite Matrices}, The Quarterly Journal of Mathematics, os-3(1):253-268, (1932).

\bibitem{Cooke 1955} Cooke, R. G.: {\it Infinite matrices and sequence spaces}, (1955), Dover Publications, Inc. New York.

\bibitem{Sivakumar 2009}  Shivakumara, P.N.  and  Sivakumar, K. C. : {\it A review of infinite matrices and their applications },Linear Algebra and its Applications, Volume \textbf{430}, Issue 4, Pages 976-998, (2009).   

\bibitem{Snyder 1947} Snyder, H.S.:  \textit{Phys. Rev.}, \textbf{72}, (1947), 38.



\end{thebibliography}
\end{document}